\documentclass[%
 reprint,
 amsmath,amssymb,
 aps,
pra]{revtex4-2}

\usepackage{graphicx}
\usepackage{dcolumn}
\usepackage{bm}
\usepackage{wrapfig}
\usepackage{subcaption}
\usepackage{amsmath}
\usepackage{mathtools}
\usepackage{xcolor}

\allowdisplaybreaks 

\begin{document}

\preprint{APS/123-QED}

\title{Transmission Loss of a Labyrinthine Acoustic Metamaterial Augmented with Multichannel Feedforward Active Noise Control}

\author{Gregory M. Hernandez}
\email{gregory.hernandez@duke.edu}
 \affiliation{%
 Institute of Sound and Vibration Research, University of Southampton (UK)
}%
 \altaffiliation[Also at ]{Duke University.}
\author{Gianluca Memoli}%
 \email{g.memoli@sussex.ac.uk}
 \affiliation{
 AURORA Project, School of Engineering and Informatics, University of Sussex (UK)
}
\altaffiliation[Also at ]{Metasonixx Ltd, Brighton (UK).}
 \author{Jordan Cheer}%
 \email{j.cheer@soton.ac.uk}
 \affiliation{%
 Institute of Sound and Vibration Research, University of Southampton (UK)
}%




\date{\today}

\begin{abstract}
Acoustic metamaterials and active noise control are two advanced noise control treatments that can typically offer performance that exceeds that of conventional passive noise control treatments. Acoustic metamaterials utilize sub-wavelength structures to realize sound field control, whilst active noise control treatments achieve control via the introduction of additional sources driven to generate a secondary sound field that interferes in a controlled way with the original, primary sound field. This paper presents an investigation into combining these two noise control techniques, to achieve enhanced noise control over a spatial region using a ``hybrid" device. In particular, conventional feedforward active noise control is combined with a labyrinthine metasurface and the increase in  performance offered by the hybrid solution is demonstrated. 

\end{abstract}

\maketitle





\begin{figure*}[t]
  \centering      
  \includegraphics[trim={0cm 0cm 0 0cm}, width=17.5cm ,clip]{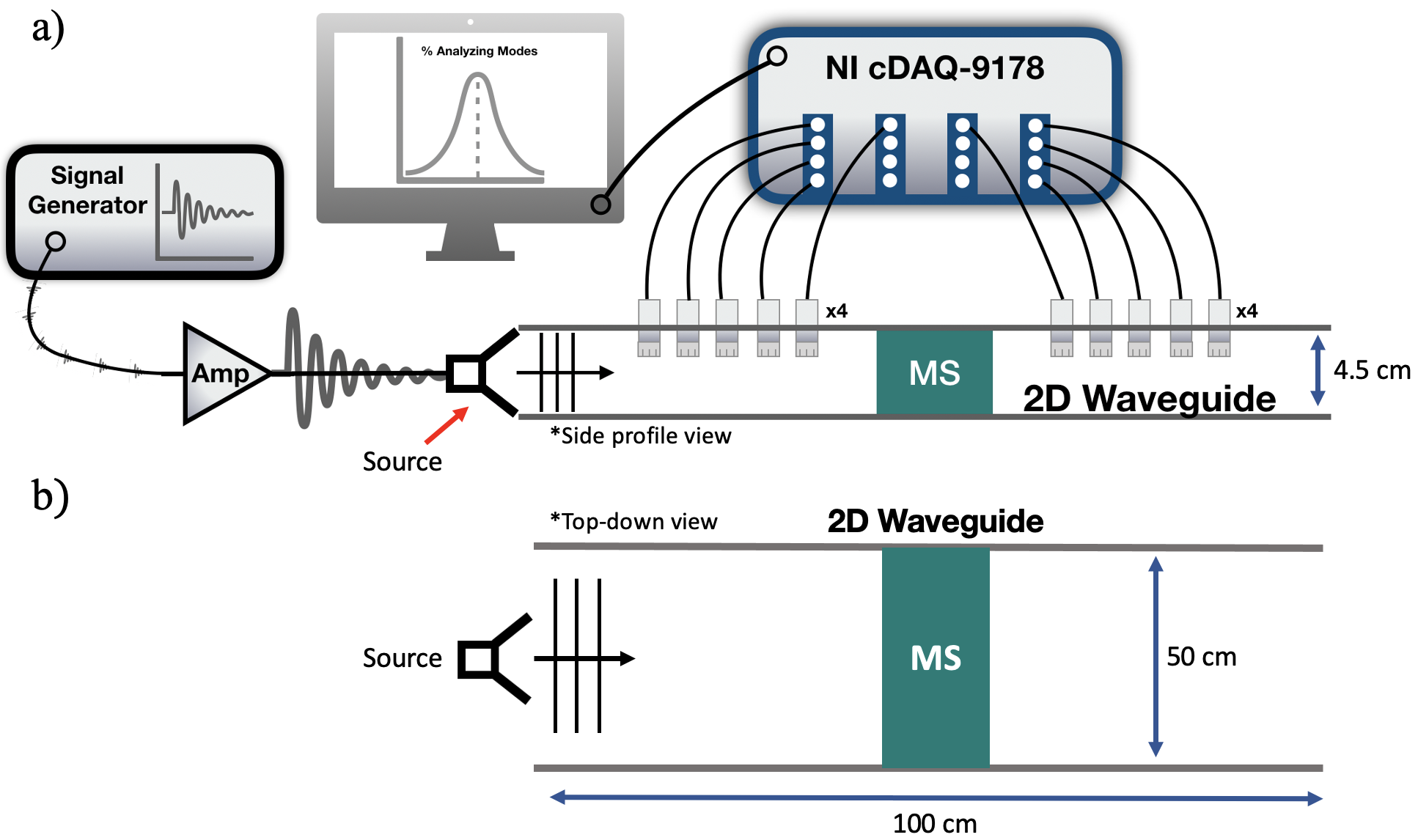}
  \caption{(a) The 2D rectangular waveguide is 100 cm x 50 cm x 4.5 cm. There are 40 microphones positioned in the upstream and downstream sections (20 each side). The signal generator sends Gaussian white noise to an amplifier that excites a loudspeaker at discrete locations. The data acquisition unit sends the sensor signals to the computer where they are stored, and processed in a computer with MATLAB. MS indicates the metamaterial position. (b) The top-down view of the waveguide.}
  \label{Figure_2}
\end{figure*}

\section{Introduction}
Industrial noise has been a part of human society since the industrial revolution, and it is now well accepted that this pollutant has a significant cost to society, due to its negative health effects \cite{WHO}. For many centuries, however, acoustical engineers had two principle methods of managing noise: mass-based solutions and acoustic absorbers \cite{Fahy}. Unless resonators are used, however, these two methods require a large spatial footprint or mass to achieve control at the lower frequencies typical of industrial settings and ventilation systems.

Active Noise Control (ANC), whose first commercial applications emerged in the 1980s \cite{ANC_Bose}, offers a solution to overcome the low frequency limitations of conventional passive noise control treatments in a variety of applications including headphones \cite{ANC_Bose}, road vehicles \cite{ANC_cars} or aircraft \cite{elliot1990flight}. There are a variety of physical aspects that limit the potential application of active noise control systems, but when the objective is to achieve sound field control over an extended spatial domain the number of secondary sources, or control loudspeakers, rapidly increases. This increases both the complexity and the cost of utilizing an active system and, therefore, their use is often limited to high performance, critical applications. 

An alternative solution to overcoming the limitations of conventional passive noise control treatments is offered by acoustic metamaterials \cite{Cummer:Review, Junfei_abs, Junfei_biani}. Typically sub-wavelength, these structures have an artificial bulk modulus and mass density that can facilitate sound field manipulation enabling applications such as beamforming  \cite{Memoli} or acoustic cloaking \cite{Cloak}. Various acoustic metamaterial designs have been proposed that are able to achieve effective sound attenuation at low frequencies, for example in \cite{Junfei_abs} an array of resonators is used to form a metasurface that achieves 99\% energy absorption at 511 Hz with a surface thickness of about $\lambda$/20th, where $\lambda$ is the acoustic wavelength; in \cite{li2016acoustic} space coiling is used to realize perfect absorption at 125~Hz with a thickness of about $\lambda$/223; and in \cite{ma2014acoustic} a membrane-type acoustic metamaterial is shown to achieve perfect absorption at 152~Hz with a thickness of around $\lambda$/133. These various metamaterial devices, however, have two key limitations: they are effective only over a limited bandwidth and can be challenging to mass-manufacture. Overcoming these two limitations has led to significant research effort in both the manufacture and application of metamaterials \cite{liao2021acoustic}. However, real-world application has remained somewhat limited, with only a few examples of metamaterials applied in practice \cite{sangiuliano2022low, pires2025novel, acoustic_bullettin}. Despite extensive development, and emerging commercial exploitation \cite{acoustic_bullettin}), the utilization of passive metamaterials for dynamic applications remains challenging.

 To address the challenge posed by dynamic applications, various researchers have proposed active acoustic metamaterial solutions for wave control \cite{Popa_active1, Popa, tan2022realisation, tan2024realisation}, however, these are typically rather complex to implement and the cost would be prohibitive for many applications. To reach a balance between performance under dynamic applications and complexity, previous research has also explored the integration of active noise control techniques with a passive Helmholtz resonator based metamaterial \cite{Cheer}. This hybrid passive-active solution demonstrated a 10~dB enhancement in transmission loss compared to either the passive or active systems operating in isolation, and achieving more than a factor of 8 increase in the bandwidth compared to the passive resonator-based metamaterial. 
  
 In this work, an alternative hybrid active-passive acoustic metamaterial is investigated, which is realized by combining a feedforward active noise control system with a static acoustic metasurface realized using the labyrinthine unit cells described in \cite{Memoli}. The objective here is to enhance the downstream attenuation performance of the metasurface by using active noise control, and thus demonstrate the benefits of a hybrid approach, providing useful insights for future development of similar hybrid metamaterials. Notably, the hybrid system is evaluated in a 2D waveguide, which goes beyond the 1D waveguide investigations used in much of the literature \cite{Cheer, Popa}. The paper is structured as follows: Section \ref{sec: experimental construction} describes the physical system, including the labyrinthine metasurface and the loudspeakers and microphones utilized to implement the active control system; Section \ref{sec: passive performance} presents the passive performance of the metasurface; Section \ref{sec: active performance} presents the performance of both the active noise control system alone and when combined in various ways with the passive metasurface to realize different hybrid systems; finally, Section \ref{sec: conclusions} presents conclusions.

\section{Experimental Construction}\label{sec: experimental construction}

The experimental apparatus is depicted in Fig.~\ref{Figure_2}a and ~\ref{Figure_2}b. The 2D waveguide - 100 cm long, 50 cm wide, and 4.5 cm tall - is constructed of 5 mm thick rigid engineering plastic (polyamide 66). The lateral walls of the waveguide are screwed to both the top and bottom plates and further secured with epoxy to the bottom plate.

\subsection{Labyrinthine Metasurface}

The considered metasurface, noted by MS in Figure \ref{Figure_2}, has been realized using the concept of ``metamaterial bricks" introduced by Memoli \emph{et al.} \cite{Memoli}. This work demonstrated that, once the main frequency of operation has been selected, most narrow-band metasurfaces can be built by reconfiguring 16 labyrinthine pre-defined metamaterial bricks, each encoding a specific phase shift between $0$ and $15/8 \pi$. In \cite{Memoli}, the metamaterial bricks have a length of $\lambda$ in the direction of propagation, which will be referred to as the ``thickness" of the device, and are $\lambda/2$ wide in the other two dimensions. However, these cells were considered too large for the current setup. At 2400~Hz the unit cells would have been 14.3~cm thick and 7.14~cm in the lateral dimensions. Therefore, the metamaterial bricks presented in \cite{memoliCHI} have been used here, which have a thickness of $\lambda/3$, a lateral dimension of $\lambda/6$, and encode a phase delay ranging from $0$ to $15/8 \pi$. 

The metasurface utilized in this experiment contains 18 unit cells, constructed of PLA plastic, giving a total width of 50~cm and a thickness of $\approx 4.76 \,$ cm. The 2D blueprints in \cite{memoliCHI} were elongated to fit the waveguide, so that the final height of the metasurface was 4.5~cm. The thickness of the metasurface was 5.7~cm (40\% or $2\lambda/5$ in air), with a physical wall construction thickness of 2~mm to ensure an effective connection with the loudspeaker array structure (see Fig.~\ref{Figure_3}b-e) and to ensure that the structure is sufficiently rigid to avoid structural effects. 

\begin{figure}
  \centering    
  \includegraphics[trim={0 0 0 0}, width=8.5cm ,clip]{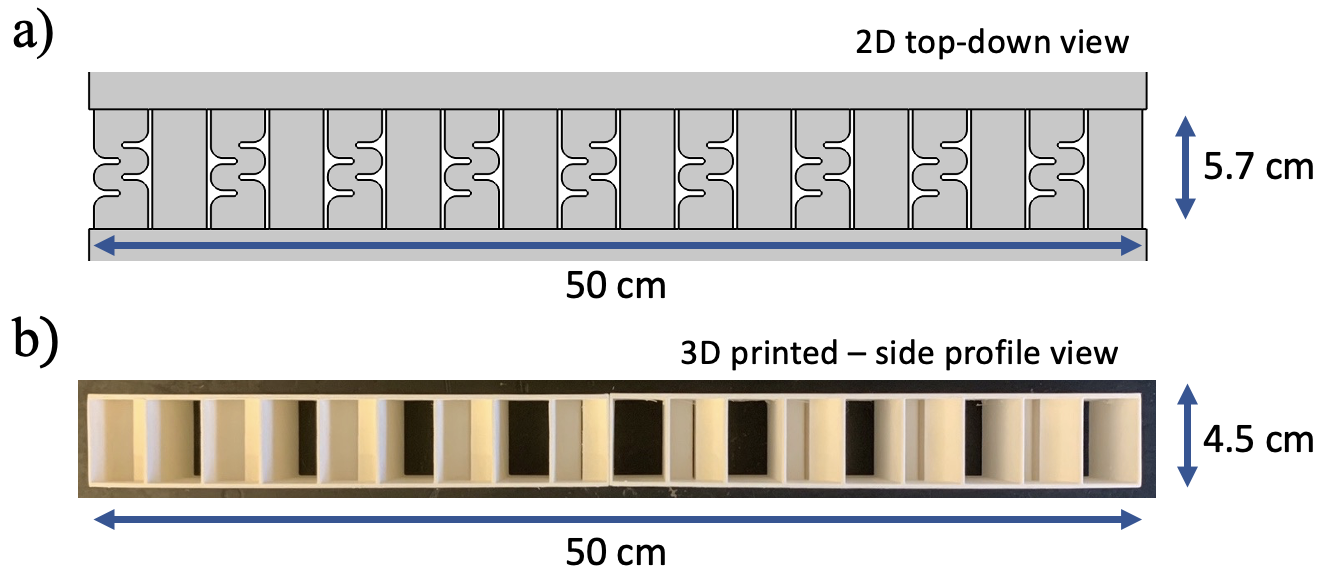}
  \caption{(a) The 2D top-down perspective of the metamaterial unit cells situated in the waveguide. (b) The 3D-printed metamaterial constructed of PLA plastic. There are a total of 18 unit cells: 9 are open units, the other 9 are labyrinth unit cells.}
  \label{Figure_4}
\end{figure}

The metasurface has been constructed from two distinct metamaterial bricks, which are arranged in an alternating pattern, as shown in Figure~\ref{Figure_4}a and~\ref{Figure_4}b. This two unit cell architecture (analogous to a dipole pair) is the basis of the metasurface design used here. At the selected frequency of 2400~Hz, one unit cell is an open channel (phase shift: $0$), through which the incident wave travels without any change in phase. The other is a meandering, or labyrinthine unit cell, designed to create a phase shift of $\pi$, so that the wave that it radiates destructively interferes with the output waveform traveling through the open unit cell at the selected frequency. For the purposes of this study, this metasurface can be considered as a passive noise-canceling device operating at a single frequency.

\subsection{Active Noise Control System}
The active noise control system consists of two principle components -- the loudspeakers used to implement control and the microphones used to monitor the sound field in the waveguide. The control system, as described in Appendix \ref{ANC System}, uses the downstream microphone signals to determine the optimal loudspeaker drive signals in order to control the downstream sound field and maximize transmission loss of the system. The following subsections describe the loudspeaker and microphone arrangements.

\subsubsection{Loudspeakers}\label{sec: loudspeakers}
\begin{figure}
  \begin{center}    
    \includegraphics[trim={0cm 0cm 0cm 0cm}, width=8.5cm ,clip]{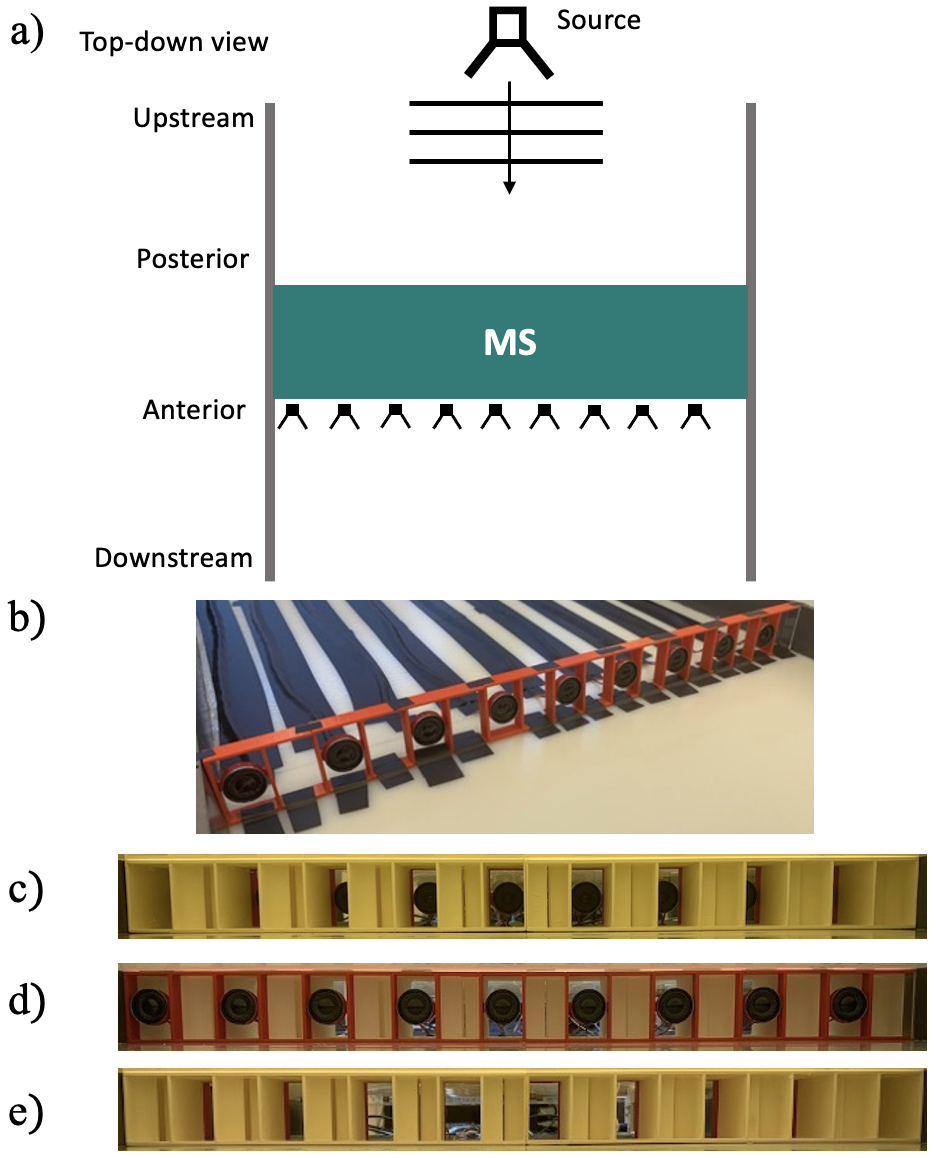}
  \end{center}
  \caption{(a) The general top-down view of the waveguide depicting the source location and positions of the metasurface and loudspeaker units for active control. (b) The loudspeaker array utilized in the active control implementation. (c) The hybrid metasurface with the loudspeakers placed in the posterior position of the open unit cells. (d) The hybrid metasurface with the loudspeakers placed in the anterior position of the open unit cells. (e) The hybrid metasurface with the loudspeakers placed in the anterior position of the labyrinth unit cells.}
  \label{Figure_3}
\end{figure}

The metasurface described in the previous section has been augmented using an array of 9 miniature loudspeakers (see Fig.~\ref{Figure_3}b) each rated as a 32 ohm, 500 mW driver with a 23 mm diameter (model: MCABS-231-RC by multicompPRO). These transducers were chosen to meet the height limitation of the waveguide and lateral opening of the metasurface unit cells, and to have a frequency response covering the bandwidth of the designed metasurface. Each transducer was housed in a 3D-printed enclosure, designed to match the channel dimensions and wall thickness of the metasurface (shown in red, in Fig.~\ref{Figure_3}b). The loudspeaker units were designed with an alternating arrangement, as shown in Fig.~\ref{Figure_3}b, to match the unit cell arrangement of the metasurface. Specifically, three loudspeaker configurations have been investigated: firstly, with the loudspeakers upstream, or posterior, of the open unit cells (Fig.~\ref{Figure_3}c); secondly, with the loudspeakers downstream, or anterior, of the open unit cells (Fig.~\ref{Figure_3}d); and finally, with the loudspeakers upstream, or posterior, of the labyrinthine unit cells (Fig.~\ref{Figure_3}e). The loudspeakers could not be placed in front of the labyrinth cells of the metasurface due to their narrow geometry.

\subsubsection{Microphones}
The minimum number of microphones to be used for active noise control has been determined using the modal decomposition approach by Zhang \emph{et al.} \cite{Zhang}. As described in Appendix~\ref{Modal Decomp}, this method allows the propagating modes within a 2D-waveguide to be detected and is therefore much simpler than the raster scanning method employed in other works \cite{Popa, Junfei_abs}.  White noise is utilized to excite the waveguide, thus allowing for a large bandwidth to be analyzed. The method proposed in \cite{Zhang} also allows the transmission and reflection coefficients for all modes within the considered frequency range of the experiment to be obtained. 

As described in Appendix~\ref{Modal Decomp}, a total of 40 sensors are needed to ensure that the modal matrix is overdetermined. Therefore, 20 electret microphones (PCB Piezotronics, Model 130F20) were placed upstream of the metamaterial, and 20 located downstream. The general location of these sensors is depicted in Fig.~\ref{Figure_2}a. The microphones are connected to a National Instruments CompactDAQ cDAQ-9178, depicted above the waveguide in Fig.~\ref{Figure_2}a, which sends the time domain microphone signals to MATLAB. The source located at the end of the waveguide, also shown in Fig.~\ref{Figure_2} outputs Gaussian white noise with a standard deviation of 1 Volt, driven via a Wondom Class D Audio Amplifier (Sure Electronics AA-AB32155). Note that in Fig.~\ref{Figure_2}a,b MS indicates the metasurface location (or any other device being measured).

\section{Passive Performance}\label{sec: passive performance}
\begin{figure*}[t]
   \centering    
   \includegraphics[trim={0cm 0cm 0 0cm}, width=14cm, clip]{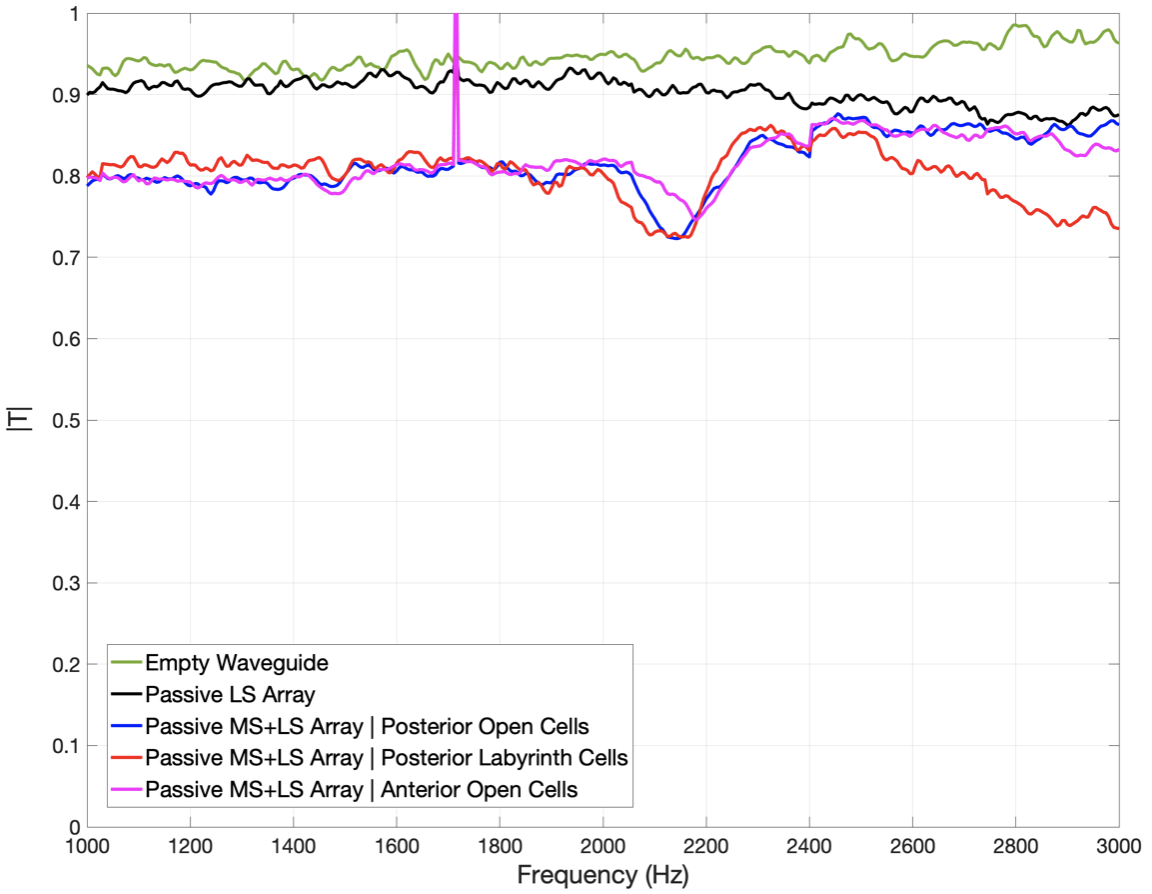}
   \caption{Passive measurements. Transmission of the 0\textsuperscript{th} propagating mode within the waveguide. Passive attenuation for an empty waveguide (green), standalone loudspeaker array (black), metasurface and loudspeaker hybrid system: loudspeakers behind open cells (blue), loudspeakers behind labyrinth cells (red), loudspeakers in front of open cells (pink). }
   \label{Figure_7}
 \end{figure*}

This section presents measurements of the passive attenuation provided by both the undriven or passive loudspeaker array within the waveguide and the hybrid labyrinthine metasurfaces. The measured passive performance is evaluated in terms of the transmission of the  0\textsuperscript{th} propagating mode. Fig.~\ref{Figure_7} shows the transmission coefficient for the empty waveguide, the passive loudspeaker array and the passive metasurface with the loudspeakers located in the possible configurations described in Section \ref{sec: loudspeakers}. It is important to first note from the results presented in Fig.~\ref{Figure_7} that a numerical error arises at 1715~Hz; this is due to the modal decomposition method in obtaining the plane wave propagation (see Appendix~\ref{Modal Decomp}), and can be attributed to the conditioning of the modal matrix derived by Zhang {\sl et al.} \cite{Zhang}. Although the conditioning of this problem may be improved via regularization techniques, since the numerical error occurs at a frequency that is below the range of interest for the designed metasurface, it  will not be considered further. 

The results presented in Fig.~\ref{Figure_7} reveal that the transmission of the empty waveguide is not exactly unitary, which can be attributed to the thermo-viscous losses and the damping within the waveguide itself. It can also be seen from Fig.~\ref{Figure_7} that the insertion of the passive loudspeaker array introduces a small decrease in the transmission. Specifically, at the frequencies between 1~kHz and 2~kHz, the transmission of the 0\textsuperscript{th} propagating mode is also 90\% with the loudspeaker array, and above 2~kHz the transmission decreases to 85\% as the frequency approaches 3~kHz. This can be related to the absorption being provided by the passive components of the loudspeaker array (i.e. the materials of the loudspeaker and the plastic housing) and, at higher frequencies, to reflection from the loudspeakers as the wavelength decreases. 

Fig.~\ref{Figure_7} also shows the passive transmission performance of the metasurface with the loudspeaker array in the three considered configurations. Within the 1000-2000~Hz bandwidth, the introduction of the metasurface results in nearly 10\% more of the incident wave being reflected and/or absorbed compared to the passive transmission of the loudspeaker array alone. A further dip in the transmission can be observed in Fig.~\ref{Figure_7} for the three passive metasurface-loudspeaker configurations between around 2000 and 2250~Hz, which defines the operating bandwidth of the passive metasurface. For the two configurations with the loudspeakers in the posterior position, the bandwidth of the metasurface is relatively consistent. However, there is a small shift in the resonant peak -- from 2150~Hz to 2180~Hz -- when comparing the two posterior responses (blue and red) to the anterior response (pink). Additionally, the level and bandwidth of transmission loss with loudspeakers in front of the open cells is not as significant as in the two posterior loudspeaker variations. At frequencies above the operational bandwidth of the passive hybrid metasurface, the transmission loss depends on the relative position of the loudspeakers: when the loudspeakers are upstream of the labyrinth, the transmission decreases with frequency, reaching around 70\% at 3 kHz;  when the loudspeakers are either upstream or downstream of the open unit cells, the transmission stays constant at around 80\%. Note that the resonance for either posterior hybrid metasurface configuration will be referenced to 2150 Hz moving forward.

It is worth discussing the difference between the measured (2150~Hz) and designed (2400~Hz) operational frequency of the passive metasurface. This can be related to the additional path length introduced in the experimental realization due to the loudspeaker array, which decreases the frequency at which effective interference between the open and labyrinth unit cells occurs. To support this hypothesis, 2D and 3D simulations were performed (using the Acoustic Module of COMSOL Multiphysics) to explore the effect of the additional thickness introduced by the inclusion of the loudspeaker array. The simulation results, reported in Appendix \ref{Sims}, show a similar decrease in the resonance frequency of around 400~Hz when the loudspeaker array is introduced either in front or behind the metasurface. The difference between the simulated and measured shift in frequency can be attributed to the approximations used in the simulations (e.g. the loudspeakers were modeled as hard objects) and to the higher-order physics interaction between the electroacoustic devices and the metasurface.

\section{Active Performance}\label{sec: active performance}
Having demonstrated the passive performance of the hybrid metasurfaces in the previous section, this section presents an investigation into the active performance of these proposed systems. Since there was no significant difference in the transmission of the 0\textsuperscript{th} propagating mode with the loudspeaker array positioned either posterior or anterior to the open unit cells of the metasurface, the hybrid performance is explored here only for the two cases where the loudspeakers are posterior to the open  or the labyrinth unit cells. Additionally, to provide context to the additional performance offer by the proposed hybrid metasurface, the active performance provided by the loudspeaker array in isolation is also analyzed.

In all three active cases, control is realized using an optimal multichannel feedforward active control system and simulated offline using the responses measured using the experimental system. This approach allows the physical limits on control performance to be investigated, without introducing the complexities of real-time implementation which are well understood \cite{Elliott2001SignalControl}. The assumed feedforward control strategy, which is introduced in Appendix \ref{ANC System}, calculates the signals required to drive the array of loudspeakers to minimize the sum of the squared pressures measured at the array of downstream microphones. The active control algorithm also includes a constraint on the control effort, which is defined as the sum of the squared magnitudes of the control signals driving the loudspeakers, as given by equation~\ref{equationB_3}. It is necessary to constrain the control effort in practice to avoid over-driving the loudspeakers, but also to make the controller robust to real-world uncertainties. A control effort constraint is imposed here so that the control signals driving the loudspeakers are realizable in practice, but also to ensure that the required control effort is consistent between the three considered active systems. 

\begin{figure*}
  \centering
  \includegraphics[width=14cm, clip]{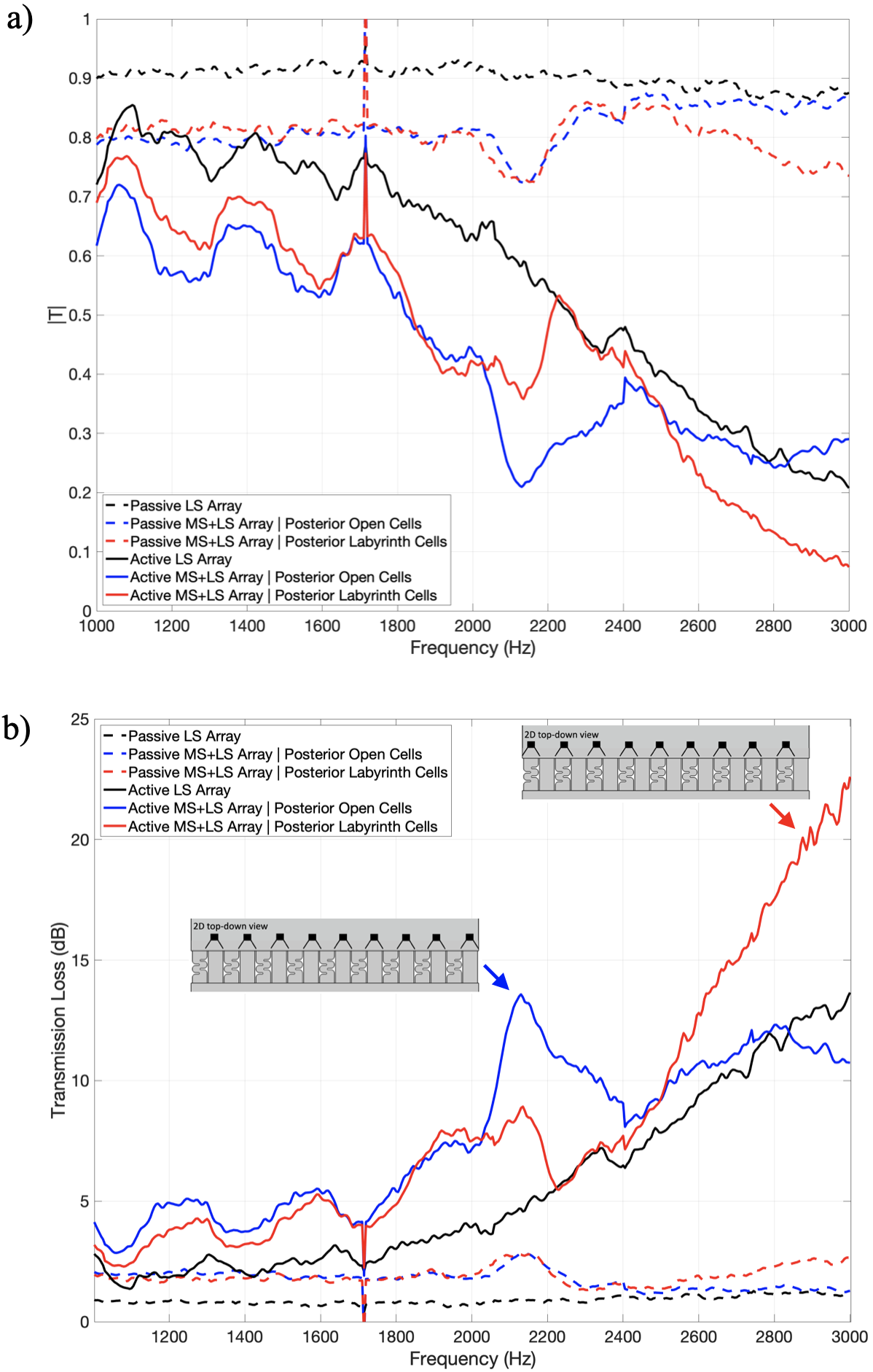}
    \caption{Transmission (a) and transmission loss (b) of the 0\textsuperscript{th} propagating mode within the waveguide. Three passive cases are presented without active control---loudspeaker array (dashed black), hybrid metasurface with loudspeakers: behind labyrinth cells (dashed red), behind open unit cells (dashed blue). The analogous offline active control applications are also given---loudspeaker array (hard black), hybrid metasurface with loudspeakers: behind labyrinth cells (hard red), behind open unit cells (hard blue). Overlaid in (b) are the 2D geometric views of the two hybrid systems analyzed for active control. }
    \label{Figure_89}
\end{figure*}

Fig.~\ref{Figure_89}a,b presents the transmission and transmission loss respectively of the three passive systems previously presented in Section \ref{sec: passive performance} along with the three active systems, including the two hybrid metasurfaces, and the standalone active loudspeaker array. Transmission loss is defined as 
\begin{equation}
    TL = 20\log_{10}(1/|T|) \:,
\label{equation1}
\end{equation}
\noindent
where $|T|$ is the magnitude of the transmission coefficient.

Firstly, the active control algorithm was applied to the standalone loudspeaker array, placed in the posterior position relative to the metasurface (although the passive metasurface itself has been removed). The sold black line in Fig.~\ref{Figure_89}a,b shows the performance of this active system and comparing this to the passive loudspeaker array shows that a significant reduction in the transmission is achieved by the active system, particularly at frequencies above 2000~Hz. This slightly curious result from the perspective of active control, which is generally more effective at lower frequencies, may be related to the fact that only the 0\textsuperscript{th} order propagating mode is being considered here. Nevertheless, it is clear from these results that the active system using the loudspeaker array in isolation does exhibit a transmission loss that increases between 1000~Hz and 3000~Hz by around 11~dB, compared to a 1 dB increase for the passive loudspeaker array (dashed black line). It is also worth noting that at the 2150~Hz resonance of the hybrid metasurfaces, the transmission loss for the active loudspeaker array is 4~dB, which is an increase of around 3~dB compared to the passive attenuation provided by the loudspeaker array at this frequency.

Fig.~\ref{Figure_89} also shows the performance of the two active hybrid metasurfaces, with the loudspeaker posterior to either the open or labyrinth unit cells (solid red and blue lines respectively). In both cases, the hybrid metasurfaces significantly outperform both their respective passive responses (dashed red and blue lines) and the purely active system. An important note is that the two hybrid metasurfaces do not achieve the same active performance, even at frequencies where the passive performance is quite similar, which highlights a difference in the interaction between the active and passive components of the system. 

The performance of the two hybrid metasurfaces can be effectively discussed considering three distinct frequency bands: below the passive metasurface interference band ($f<$2 kHz); within the passive metasurface interference band (2 kHz$f<$2.25 kHz); and above the passive metasurface interference band ($f>$2.25 kHz). At frequencies below the passive metasurface interference band, the two hybrid metasurfaces achieve quite consistent active performance, with an average level of around 6~dB. The hybrid systems also provide a significant performance advantage compared to the purely active system over this lower frequency bandwidth, with an increase in the transmission loss of up to 3~dB. At frequencies around the passive metasurface interference band, the hybrid system with the loudspeakers posterior to the open unit cells achieves the highest performance, with a transmission loss of 13~dB; while the hybrid metasurface with the loudspeakers positioned posterior to the labyrinth unit cells achieves a maximum transmission loss in this bandwidth of 9~dB. Finally, at frequencies above the passive metasurface interference band, the hybrid system with the loudspeakers posterior to the open unit cells achieves performance largely consistent with the purely active system, with a maximum transmission loss of 12~dB. In the same frquency band, the transmission loss achieved by the metasurface with the loudspeakers posterior to the labyrinth unit cells increases at a greater rate than the other active systems, providing a transmission loss up to 11~dB greater than the other hybrid metasurface configuration. 

The results presented for both hybrid metasurface configurations demonstrate a significant performance advantage compared to traditional active noise control, both in the frequency band for which the passive metasurface is designed and beyond. However, the differences in performance between the two hybrid metasurfaces highlights differences in the coupling between the loudspeakers and the passive metasurface and it is insightful to discuss the potential underlying physical mechanisms.

\begin{figure*}[t]
  \centering      
  \begin{subfigure}{.5\textwidth}
  \centering
  \includegraphics[width=.85\linewidth]{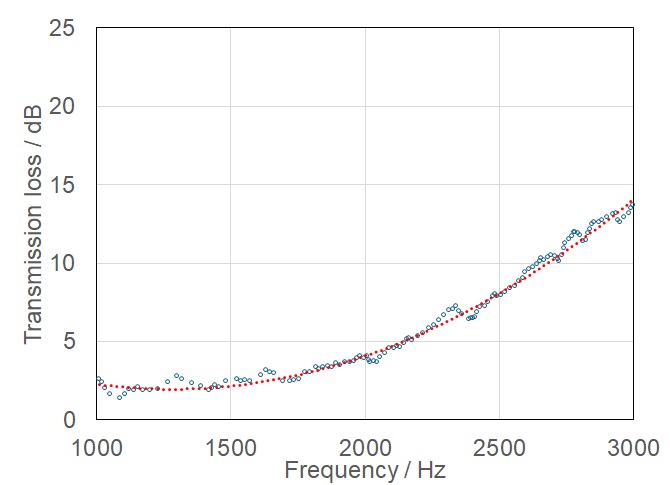}
  \caption{Background data, fitted with a 2nd degree polynomial.}
\end{subfigure}%
\begin{subfigure}{.5\textwidth}
  \centering
  \includegraphics[width=.9\linewidth]{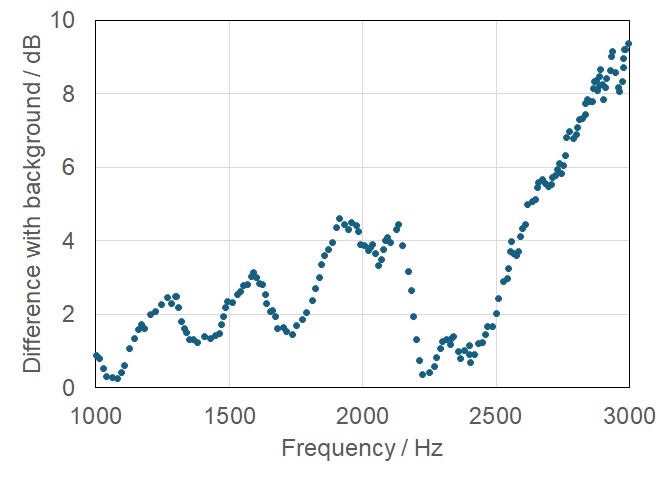}
  \caption{Signal data: loudspeakers behind the labyrinthine cells.}
\end{subfigure}%
\\
\begin{subfigure}{.5\textwidth}
  \centering
  \includegraphics[width=.9\linewidth]{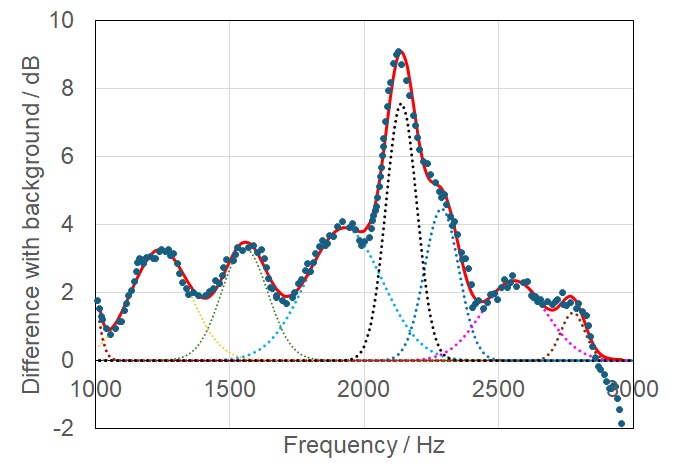}
  \caption{Multi-peak fit, open cells.}
\end{subfigure}%
\begin{subfigure}{.5\textwidth}
  \centering
  \includegraphics[width=.9\linewidth]{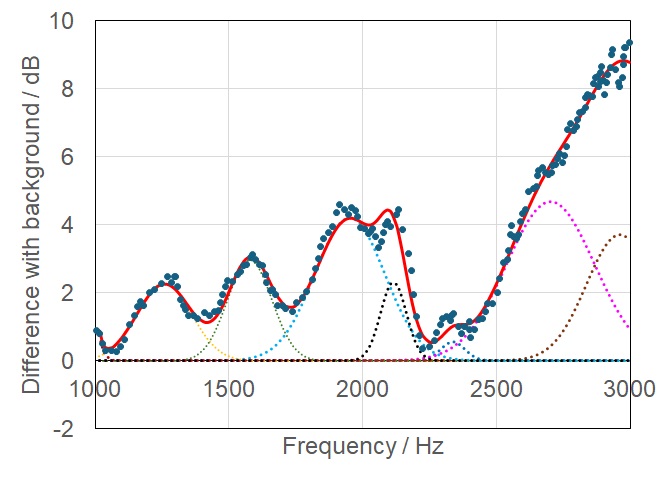}
  \caption{Multi-peak fit, labyrinthine cells.}
\end{subfigure}%
  \caption{A schematic of the spectroscopy analysis: (a) background fit; (b) signal; (c) multi-peak fit in the case with loudspeakers behind the open cells; (d) multi-peak fit in the case with loudspeakers behind the labyrinthine cells.}
  \label{Figure_spectroscopy}
    
\end{figure*}

To investigate the nature of the coupling, in this study we analyzed the data using the techniques of optical/impedance spectroscopy \cite{Impedance2005}: 
\begin{enumerate}
    \item We assumed that the “active control only” is the background performance i.e. that the hybrid system performs at least as that. We then fitted the background TL with a polynomial, finding that a 2nd degree polynomial was sufficient i.e. $TL = 4e-6 \cdot f^2 -0.0105\cdot f + 8.71 $ where $f$ is the frequency in Hz, as shown in Figure \ref{Figure_spectroscopy}a.
    \item We subtracted the fitted background from the TL curves with the speakers in the two configurations -- see e.g. Figure \ref{Figure_spectroscopy}b for the case with the loudspeakers behind the labyrinthine cells.
    \item We used the Peak Finder function in OriginPro (OriginLab, version 2025b) to fit the obtained curves with multiple Gaussian peaks. The algorithm found 13 peaks, which were examined for relevance: eliminating those with too much uncertainty, we reduced the number of peaks to 7. The resulting fits can be found in Figures \ref{Figure_spectroscopy}c,d and the coefficients in \ref{Tab_spectroscopy}. 
\end{enumerate}
\begin{table*}
\centering
\caption{Coefficients of the multi-peak fits in Figure \ref{Figure_spectroscopy}. Uncertainty (not reported) were approximately 10\% of the values in the table.}
\label{Tab_spectroscopy}
\begin{subtable}{\textwidth}
\begin{tabular}{|l|c|c|c|c|c|c|c|c|c|}
\hline
 & Peak 0 &	Peak 1 &	Peak 2 &	Peak 3 &	Peak 4 &	Peak 5 &	Peak 6 &	Peak 7 &	Peak 8\\
\hline
Amplitude / dB &	7	&3	&3	&4	&8	&4	&2	&1	& n.a.\\
Frequency / Hz	& 940	& 1242 & 	1556 & 	1926 &	2139 &	2289	& 2558	& 2778 & n.a.\\	
FWMH / Hz &	56 &	157  &	121	&200	& 82	& 90	& 169 &	63	& n.a.\\
\hline
\end{tabular}
\caption{Fitting values for the ``open cells" case.}
\end{subtable}
\begin{subtable}{\textwidth}
\begin{tabular}{|l|c|c|c|c|c|c|c|c|c|}
\hline
 & Peak 0 &	Peak 1 &	Peak 2 &	Peak 3 &	Peak 4 &	Peak 5 &	Peak 6 &	Peak 7 &	Peak 8\\
\hline
Amplitude / dB	& 7	& 2	& 3	& 4	& 2	& 1	& 5	& 4	& 6\\
Frequency / Hz	& 940	& 1258	& 1578	& 1951	& 2114	& 2335 & 	2701	& 2959	& 3164\\
FWMH / Hz	& 51	& 146	& 117	& 185	& 68	& 63	& 233	& 176	& 314\\
\hline
\end{tabular}
\caption{Fitting values for the ``labyrinthine cells" case.}
\end{subtable}
\end{table*}
\quad

As shown in Table \ref{Tab_spectroscopy}, the spectra with the loudspeakers in the two cases (i.e. behind labyrinthine \& behind open cells) show two components: 
\begin{enumerate} 
\item a single peak at $2125 \pm 20$ Hz, with an average FWMH (Full Width at Maximum Height) of 70 Hz, which we attributed to the metasurface.
\item a periodic pattern, with peaks separated by $285 \pm 60$ Hz, which we attributed to resonances in the system (e.g. to the dimensions of the waveguide).
\end{enumerate}
The efficiency of absorbing energy through the peak at 2125 Hz depends on the configuration: transmission loss at this frequency is more effective when the loudspeakers are behind the open cells. When the loudspeakers go behind the labyrinthine cells, in fact, the energy transfer changes: the energy initially transferred to the single peak now gets distributed to peaks at higher frequencies, so much that an additional peak of the periodic system (i.e. at 3164 Hz) was needed to complete the fit. The description of our hybrid metasurface in terms of the linear combination of two systems, however, has some limitations: 
\begin{enumerate} 
\item If the periodic pattern is a feature of the geometry, it should also appear in the ``active control only" case, but it was not possible to isolate it there. The energy transfer phenomenon is only visible when the metasurface is present, and this observation justifies the use of the ``active control only" data as ``background". 
\item In the ``open cells" case, the difference between the signal and the background is negative above 2860 Hz. Once again, the periodic pattern should appear, if present.   
\end{enumerate}
These two observations suggest a different physics.
The coupling between a single-line/narrow-band resonator (i.e. the metasurface) and a system capable of finding solutions at multiple frequencies, almost continuously (i.e. the active noise control system) 
has been described by Fano in 1961 \cite{Fano61}, in the case of two electronic configurations of a He atom. As shown in more recent studies \cite{Zeb2022}, however, this type of interaction can lead to two regimes: one of weak coupling, where the response appears like a single peak over an existing trend, and one of strong coupling, where the single narrowband peak splits and two peaks appear, separated in frequency and introducing a change in the trend. Considering the hybrid metasurface with the loudspeakers posterior to the open cells, the results are characteristic of a weak-coupling regime, with the trend outside the narrowband metasurface region ($\sim$2150~Hz) being similar to the one obtained with the purely active system, only with an enhancement of its periodic response. Instead, in the case where the loudspeakers are posterior to the labyrinthine unit cells, the performance appears to be the one of a strong-coupling regime, with the presence of two peaks. One peak has been shifted to a frequency lower than 2150 Hz and the other above 3000 Hz. In this explanation, the periodic pattern would decrease in intensity as the frequency increases. A deeper understanding of how to pilot these couplings (which in atomic physics are called ``polaritons" \cite{Zeb2022}) will be the subject of future studies.

\section{Conclusions}\label{sec: conclusions}
In this work, we presented an investigation into the attenuation of wave transmission using ``hybrid" metasurfaces, realized by combining a multichannel feedforward active control system with a passive labyrinthine acoustic metasurface. 
Modal decomposition was utilized to analyze the transmission of the 0\textsuperscript{th} mode propagating in a rectangular waveguide, for the cases of the passive hybrid metasurfaces, traditional active noise control and two hybrid metasurface configurations. Both active hybrid metasurfaces outperformed their respective passive responses and the purely active system, demonstrating the advantages offered by hybrid systems for broadband noise management applications.
The presented results show evidence that it is possible to pilot the physical coupling between active control and metamaterials simply by mechanically positioning the loudspeakers in different positions relative to a metasurface. Like in passive metamaterials, geometrical and design choices may result in performance changes, opening a new design space for active metamaterials.

This research has shown that the hybrid metasurfaces offer a synergistic performance advantage, with the performance exceeding that offered by the simple linear summation of the passive metasurface and the traditional active control system over different frequency ranges depending on the selected configuration. As well as investigating real-time implementation of the proposed hybrid metasurfaces, it will also be important to further optimize the metasurface to maximise the transmission loss. For example, exploring the potential performance when loudspeakers are located behind both the labyrinth and open unit cells may combine the synergistic performance advantages offered by the two hybrid configurations investigated in this work. Additionally, further work is required to investigate the coupling between the active control system and the metasurface, to further explore the concept of ``polaritons" within this acoustic metasurface context, potentially allowing greater leverage of the strong attenuation outside the resonant bandwidth of the active metasurface.

\section{Acknowledgements}
Gregory Hernandez would like to thank the US-UK Fulbright Commission for providing him the opportunity to study abroad and propose this research at the ISVR. Greg would also like to thank Ze Zhang for their help in assisting with the modal decomposition analysis. Gianluca Memoli acknowledges funding through his collaboration with Metasonixx Ltd. Jordan Cheer was partially supported by the Department of Science, Innovation and Technology (DSIT) Royal Academy of Engineering under the Research Chairs and Senior Research Fellowships programme.

\appendix
\renewcommand{\thefigure}{A\arabic{figure}}
\setcounter{figure}{0}

\section{Modal Decomposition}\label{Modal Decomp}
\begin{figure}[t]
  \centering      
  \includegraphics[trim={0cm 0cm 0cm 0cm}, width=8.5cm ,clip]{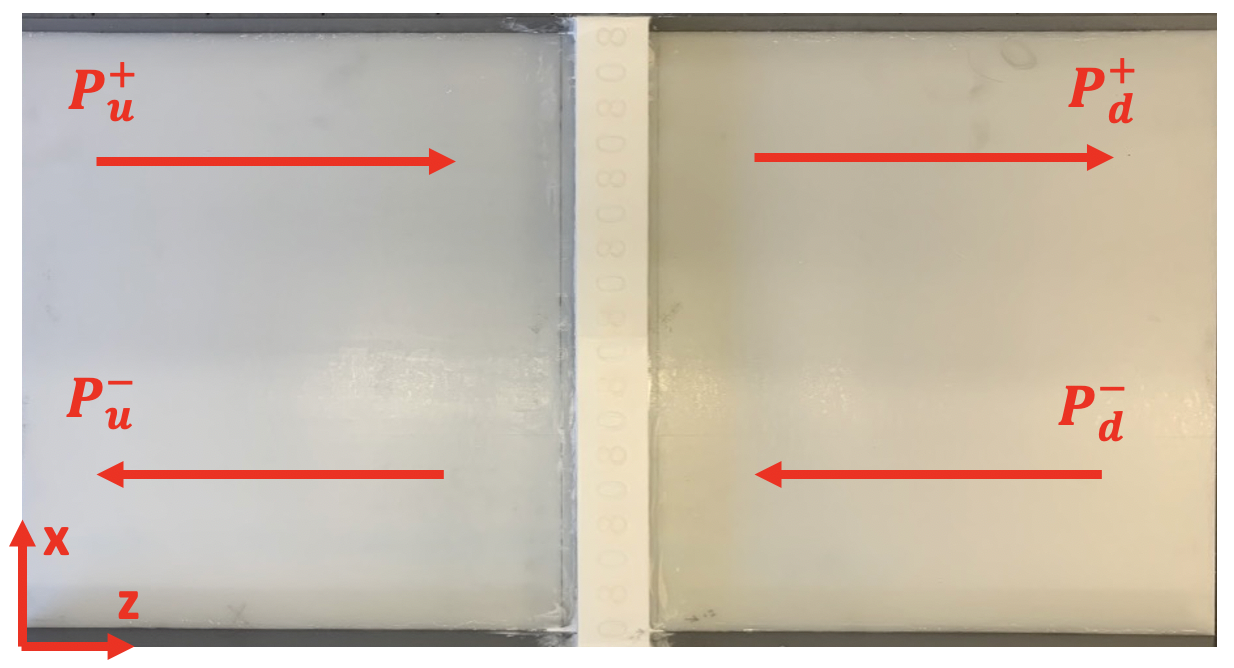}
  \caption{A schematic of the scattering matrix variables overlaid on a real image of the waveguide used in the experiment (top-down view). In the center is the metasurface.}
  \label{Figure_1}
\end{figure}

\begin{figure*}[t]
  \centering      
  \includegraphics[trim={0cm 0cm 0cm 0cm}, width=10.5cm ,clip]{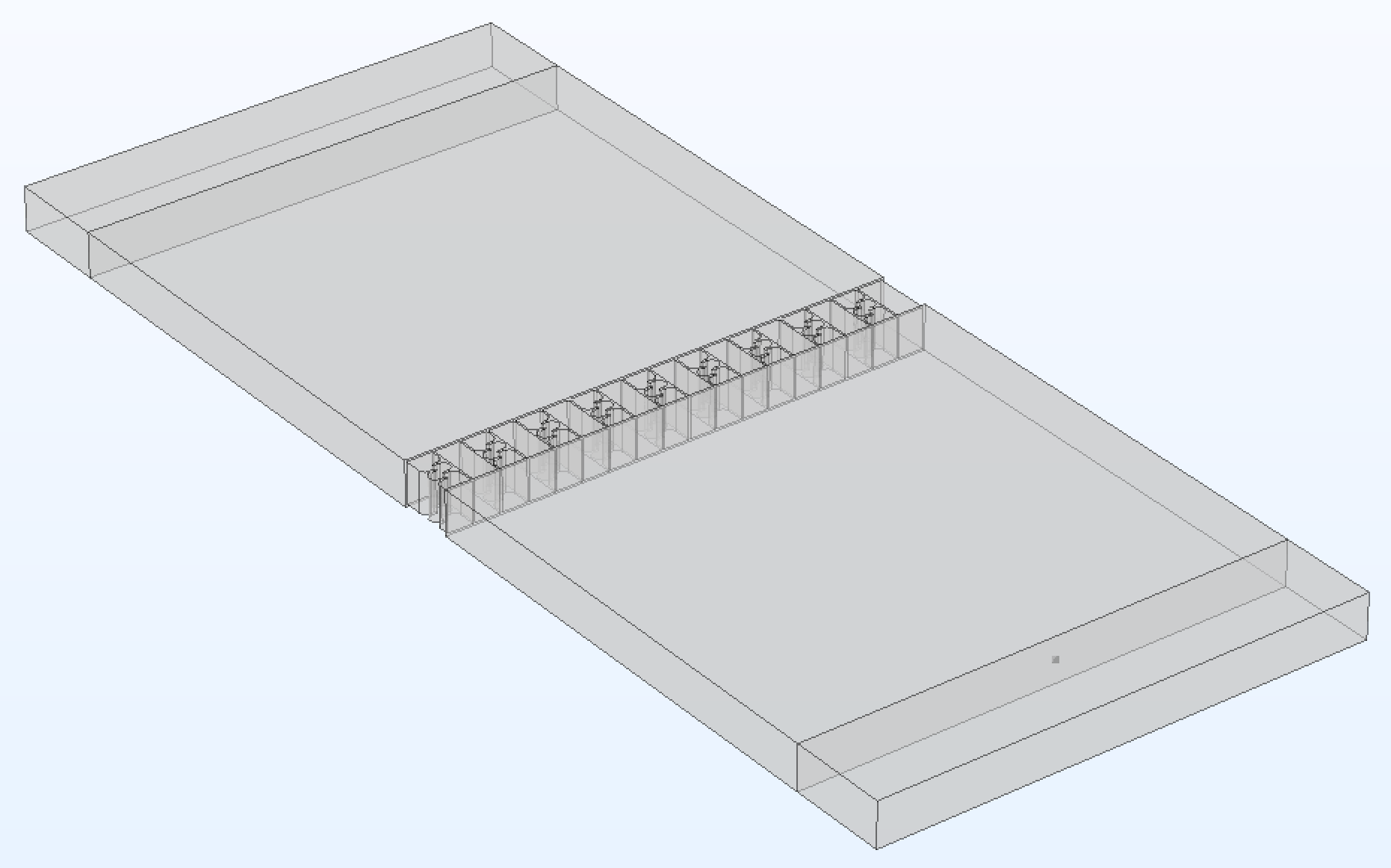}
  \caption{COMSOL Multiphysics 3D geometry experimental setup with the metasurface placed in the center of the waveguide.}
  \label{Figure_10}
\end{figure*}

This first appendix reviews the theoretical analysis necessary to obtain the modal information of the 2D-waveguide under consideration in this paper. This modal decomposition follows the derivations provided by Zhang \emph{et al.} \cite{Zhang}. 

The frequency range of interest in this paper is from 1 kHz to 3 kHz. Based on the dimensions of the waveguide (W: 500 mm, L: 1000 mm, H: 45 mm), there are 8 cut-on modes within the waveguide along the \emph{x}-dimension as depicted in Fig.~\ref{Figure_1}. The first cut-on mode in the \emph{y}-direction is at 3.81 kHz (dimension out of the page in Fig.~\ref{Figure_1}). The modes can propagate along the \emph{z}-axis and the modal matrix derived by Zhang \emph{et al.} accounts for the surface area impedance mismatch at both ends of the waveguide \cite{Zhang}. Note that the modal matrix contains the pressure pattern of the wave along the cross-section of the waveguide and the propagating wave information based on the wavenumber \cite{Zhang}. The metasurface shown in Fig.~\ref{Figure_1} splits the waveguide into two sections---an upstream and downstream portion. The pressure amplitudes and direction of the propagating modes are depicted by the red arrows in Fig.~\ref{Figure_1} where $P_u^\pm$ and $P_d^\pm$ are the pressure amplitudes of each mode---$\pm$ indicates positive or negative direction given by the coordinate system in Fig.~\ref{Figure_1}, and the subscripts $u,d$ refer to either the upstream or downstream section. Equation~\ref{equationA_1} relates the pressure amplitudes of each upstream and downstream mode to one another through the scattering matrix $\underline{S}$

\begin{equation}
\left[\frac{P_d^+}{P_u^-}\right] = \left[\frac{\underline{T^+}}{\underline{R^+}}
\frac{\underline{R^-}}{\underline{T^-}}\right]\left[\frac{P_u^+}{P_d^-}\right] = \underline{S}\left[\frac{P_u^+}{P_d^-}\right].
\label{equationA_1}
\end{equation}

\noindent
The matrices $\underline{T^\pm}$ and $\underline{R^\pm}$ represent the transmission and reflection coefficients for the different modes with positive and negative \emph{z}-direction incidences \cite{Zhang}. The modal matrix derived by Zhang \emph{et al.} requires $s = 2N$ independent measuring positions ($s$ being the number of microphones) \cite{Zhang}. The scattering matrix ($\underline{S}$) is of size $2N\mbox{x}2N$ where $N$ is the number of cut-on modes being considered \cite{Zhang}. Since there are 8 cut-on modes being considered, there must be at least 18 microphones on one side of the waveguide to capture all pressure amplitudes (this accounts for the plane wave too), and at least 18 independent measurements to obtain the scattering matrix $\underline{S}$. This experiment utilized a total of 40 microphones (20 in the upstream or downstream section of the waveguide) and 22 independent measuring positions (split evenly in the upstream and downstream section of the waveguide) to make the modal and scattering matrices overdetermined. 

The scattering matrix is further defined as 
\begin{equation}
\begin{split}
\underline{S}=\begin{bmatrix}
P_{d,0,1}^+ & \dots & P_{d,0,2N}^+ \\
\vdots && \vdots \\
P_{d,N-1,1}^+ & \dots & P_{d,N-1,2N}^+ \\
P_{u,0,1}^- & \dots & P_{u,0,2N}^- \\
\vdots && \vdots \\
P_{u,N-1,1}^- & \dots & P_{u,N-1,2N}^- \\
\end{bmatrix}
\\%
\begin{bmatrix}
P_{u,0,1}^+ & \dots & P_{u,0,2N}^+ \\
\vdots && \vdots \\
P_{u,N-1,1}^+ & \dots & P_{u,N-1,2N}^+ \\
P_{d,0,1}^- & \dots & P_{d,0,2N}^- \\
\vdots && \vdots \\
P_{d,N-1,1}^- & \dots & P_{d,N-1,2N}^- \\
\end{bmatrix}^{-1},
\label{equationA_2}
\end{split}
\end{equation}

\noindent
where in $P_{*,k,l}^\pm$, \emph{k} represents the \emph{k}th mode, and \emph{l} denotes the \emph{l}th measurement \cite{Zhang}. The elements of equation~\ref{equationA_2} depend upon the inversion of the modal matrix defined by Zhang \emph{et al.} \cite{Zhang}. To ensure the accuracy of the scattering matrix the conditioning number of the modal matrix must be small \cite{Zhang}. The modal matrix depends upon microphone position, which were optimized utilizing \textbf{fmincon} in MATLAB. The conditioning number of the modal matrix was reduced from 2.08 to 0.87 which strongly reduced the cut-on mode information as seen with the fifth cut-on mode (1715 Hz) in Figs.~\ref{Figure_7} and ~\ref{Figure_89}.

From equation~\ref{equationA_2} the absorption, transmission, and reflection of each mode can be given by

\begin{equation}
\alpha_m = 1 - \lvert\underline{S}_{(1:N,m)}\rvert^2 - \lvert\underline{S}_{((N+1):2N,m)}\rvert^2\:,
\label{equationA_3}
\end{equation}

\noindent
where $\underline{S}_{(k,l)}$ represents the \emph{k}th row and \emph{l}th column of the scattering matrix $\underline{S}$ \cite{Zhang}. The transmission and reflection coefficients to various modes from the \emph{m}th mode incidence are given as $\underline{S}_{(1:N,m)}$ and $\underline{S}_{((N+1):2N,m)}$ respectively. Making use of the term $\underline{S}_{(1:N,m)}$ from equation~\ref{equationA_3}, the transmission of the 0\textsuperscript{th} propagating mode in the positive direction, or plane wave, within the waveguide can be determined. This is the resulting transmission seen in Figs.~\ref{Figure_7} and ~\ref{Figure_89}.


\begin{figure*}[t]
  \centering      
  \includegraphics[trim={2cm 3cm 0 2cm}, width=12.5cm ,clip]{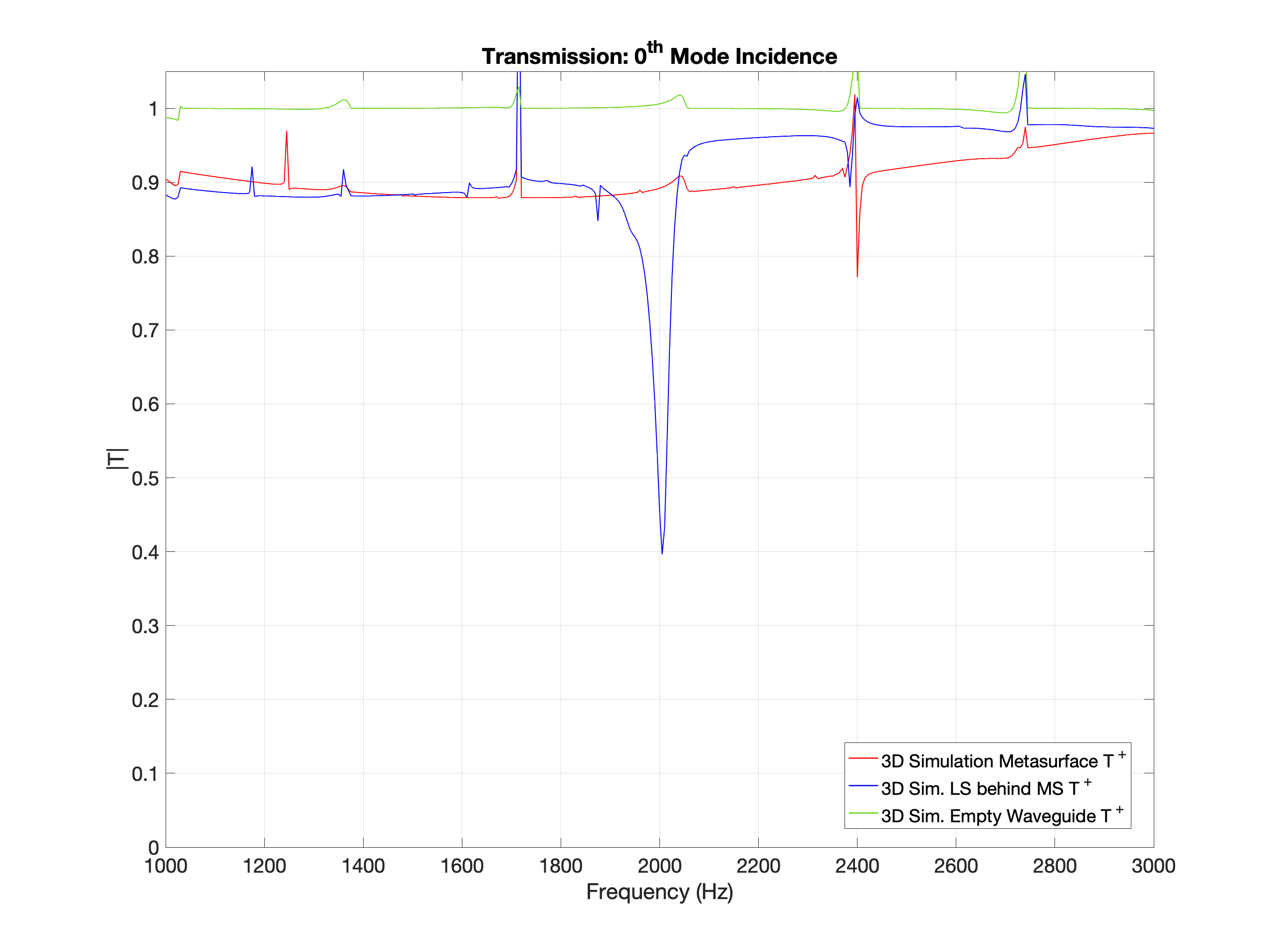}
  \caption{The 3D simulation passive transmission results for the empty waveguide (green), metasurface (red) and loudspeaker-metasurface combination with the loudspeakers behind the open unit cells (blue).}
  \label{Figure_11}
\end{figure*}

\section{Multichannel Feedforward Active Control System} \label{ANC System}
This second section introduces the multichannel feedforward active control algorithm that was applied to the loudspeaker-metasurface combination to create an active metasurface. A feedforward system, unlike a feedback one, senses a disturbance upstream from the control system and this disturbance is utilized as a reference signal for control downstream of the system. The downstream portion of the system contains an error sensor where the signal is controlled. The following equation generalizes a multichannel feedforward system

\begin{equation}
\textbf{e} = \textbf{d} + \textbf{G}\textbf{u},
\label{equationB_1}
\end{equation}

\noindent
where \textbf{e} is the vector of error sensors, \textbf{d} is the vector of disturbance signals, \textbf{G} is the matrix of complex plant responses, and \textbf{u} is the vector of control signals applied to the loudspeakers \cite{Elliott}. The plant response, \textbf{G}, represents the transfer function between the error sensors and individual loudspeakers. 

To determine optimal control signals, it is necessary to define a cost function to be minimized, which in active noise control is typically provided by the summation of the modulus squared error signals given as

\begin{equation}
J = \sum_{l=1}^L\lvert e_l\rvert^2 = \textbf{e}^{\mathrm{H}}\textbf{e},
\label{equationB_2}
\end{equation}

\noindent
where the power H is the Hermitian transpose, and $L$ is the number of error sensors \cite{Elliott}. In most physical systems, the loudspeakers have a power limiting factor which requires a constraint on the control effort to be introduced. The control effort is given as

\begin{equation}
P = \sum_{m=1}^M\lvert u_m\rvert^2 = \textbf{u}^{\mathrm{H}}\textbf{u},
\label{equationB_3}
\end{equation}

\noindent
which is related to the electrical power required to drive the loudspeakers, and $M$ is the number of actuators used for active control \cite{Elliott}. The cost function given by Equation~\ref{equationB_2} is then modified to include a term proportional to the control effort

\begin{equation}
J = \textbf{e}^{\mathrm{H}}\textbf{e} + \beta\textbf{u}^{\mathrm{H}}\textbf{u},
\label{equationB_4}
\end{equation}

\noindent
where $\beta$ is the regularization factor which is a positive real effort-weighting parameter \cite{Elliott}. The regularization factor is utilized to adjust the constraint on the control effort, which can limit the level of control signals to be within the limitations of the selected loudspeakers and can also improve the robustness of the system to real-world uncertainty. 

As mentioned, the objective of the active control system is to minimize the cost function defined by equation~\ref{equationB_4}. The multichannel feedforward active control system considered here is overdetermined, since it has more error sensors than control sources and the optimal vector of control signals is thus given as

\begin{equation}
\textbf{u}_{\mathrm{opt}} = -\left[\textbf{G}^{\mathrm{H}}\textbf{G} + \beta\textbf{I}\right]^{-1} \textbf{G}^{\mathrm{H}}\textbf{d},
\label{equationB_5}
\end{equation}


\noindent
where $\textbf{I}$ is the identity matrix with size $2M\mbox{x}2M$, and the matrix $\textbf{G}^{\mathrm{H}}\textbf{G}$ is assumed to be positive definite \cite{Elliott}. Note that the optimized control vector, and derivation of the aforementioned equations, are calculated based on frequency domain information. Equation~\ref{equationB_5} is applied to equation~\ref{equationB_1} to evaluate the error signal at the downstream section of the waveguide. Equation~\ref{equationB_5} is computed for each frequency (1 kHz to 3 kHz in steps of 5 Hz) for a chosen value of beta based on the control effort defined in equation~\ref{equationB_3}. Since there are multiple independent sources based on the modal decomposition approach, equation~\ref{equationB_5} is applied to each measurement, for all frequencies, with a different regularization factor. 

To evaluate a reasonable regularization factor, the control effort of each individual loudspeaker, for a given measurement, was compared to the rms voltage of the signal driving the loudspeakers. The excitation signal was a Gaussian white noise signal with a 1-Volt standard deviation and mean of zero. The control effort for each loudspeaker was compared to a value of 1 and a regularization factor was chosen so that no single loudspeaker out of the entire array would go above this defined threshold. 

\section{3D Simulation}\label{Sims}


Fig.~\ref{Figure_10} presents the geometry of the 3D simulation of the experimental setup including the metasurface located in the center of the waveguide using COMSOL Multiphysics. The maximum element size for the mesh of the domain was chosen to be $\frac{1}{10}$ of the wavelength for the highest frequency (3 kHz) under consideration. The waveguide had a perfectly matched layer (PML) appended to each end that represented the open end termination in the real experiment. All other boundaries were chosen as sound hard boundaries. The excitation source was chosen to be a point monopole source. Simulations were run in steps of 5 Hz from 1000 Hz to 3000 Hz. Pressure responses were recorded using domain point probes and were located based on the location of the microphones determined from Appendix~\ref{Modal Decomp}. Thermoviscous effects were not considered for this experiment. 

The 3D simulation transmission results for the metasurface (red) and the metasurface-loudspeaker variation with loudspeakers behind the open unit cells (blue) are presented in Fig.~\ref{Figure_11}. The most important part of this plot is the noticeable shift in the resonant frequency of either device. The difference in frequency is 400 Hz between the metasurface-loudspeaker combination and just metasurface. This is expected since the effective path length of the wave is increased by the addition of the loudspeaker unit cells. Secondly, the transmission loss capabilities increased by 40\% between the two metasurface cases with the passive metasurface-loudspeaker response outperforming the standalone metasurface transmission results.  

It is also worth noting that the 5\textsuperscript{th} cut-on mode and other higher-order mode contributions are seen in both transmission results of Fig.~\ref{Figure_11}. As mentioned in Section~\ref{sec: passive performance} the possible cause for the shift in frequency is due to the passive electroacoustic coupling of the loudspeaker and unit cells of the metasurface.

\medskip



\bibliography{UOS}

\end{document}